\documentclass{article}
\usepackage[T1]{fontenc}
\usepackage[utf8]{inputenc}
\usepackage{enumitem}
\usepackage[]{ismir} 
\usepackage{amsmath,amssymb,cite,url}
\usepackage{booktabs}
\usepackage{graphicx}
\usepackage{color}
\usepackage{microtype}

\title{Expotion: Facial Expression and Motion Control for Multimodal Music Generation}

\oneauthor
  {Fathinah Izzati$^{\ast}$\thanks{$^{\ast}$These authors contributed equally to this work.} \hspace{3em}
   Xinyue Li$^{\ast}$\footnotemark[1] \hspace{3em}
   Gus Xia}
  {Mohamed bin Zayed University of Artificial Intelligence, United Arab Emirates\\
   \texttt{\{fathinah.izzati, xinyue.li, gus.xia\}@mbzuai.ac.ae}}

\def\authorname{F. Izzati, X. Li, and G. Xia}

\begin{document}

\maketitle

\begin{abstract}
We propose \textsc{Expotion} (Facial \textbf{Exp}ression and M\textbf{otion} Control for Multimodal Music Generation), a generative model leveraging multimodal visual controls—specifically, human facial expressions and upper-body motion—as well as text prompts to produce expressive and temporally accurate music. We adopt parameter-efficient fine-tuning (PEFT) on the pretrained text-to-music generation model, enabling fine-grained adaptation to the multimodal controls using a small dataset. To ensure precise synchronization between video and music, we introduce a temporal smoothing strategy to align multiple modalities. Experiments demonstrate that integrating visual features alongside textual descriptions enhances the overall quality of generated music in terms of musicality, creativity, beat-tempo consistency, temporal alignment with the video, and text adherence, surpassing both proposed baselines and existing state-of-the-art video-to-music generation models. Additionally, we introduce a novel dataset consisting of 7 hours of synchronized video recordings capturing expressive facial and upper-body gestures aligned with corresponding music, providing significant potential for future research in multimodal and interactive music generation. Code, demo and dataset are available at \url{https://github.com/xinyueli2896/Expotion.git}
\end{abstract}

\section{Introduction}\label{sec:introduction}

Music generation models have become increasingly versatile and interactive, capable of integrating control signals from various modalities, including audio, text, and symbolic representations (such as MIDI or musical scores). These conditions act as control to guide the model toward producing more precise, and targeted outputs aligned with user expectations. Although current text-to-music generation models can produce impressive musical quality, they often lack the fine-grained temporal control mechanisms and expressivity necessary to flexibly adapt to a wide range of real-world scenarios.

\begin{figure}
    \centering
    \includegraphics[width=1\linewidth]{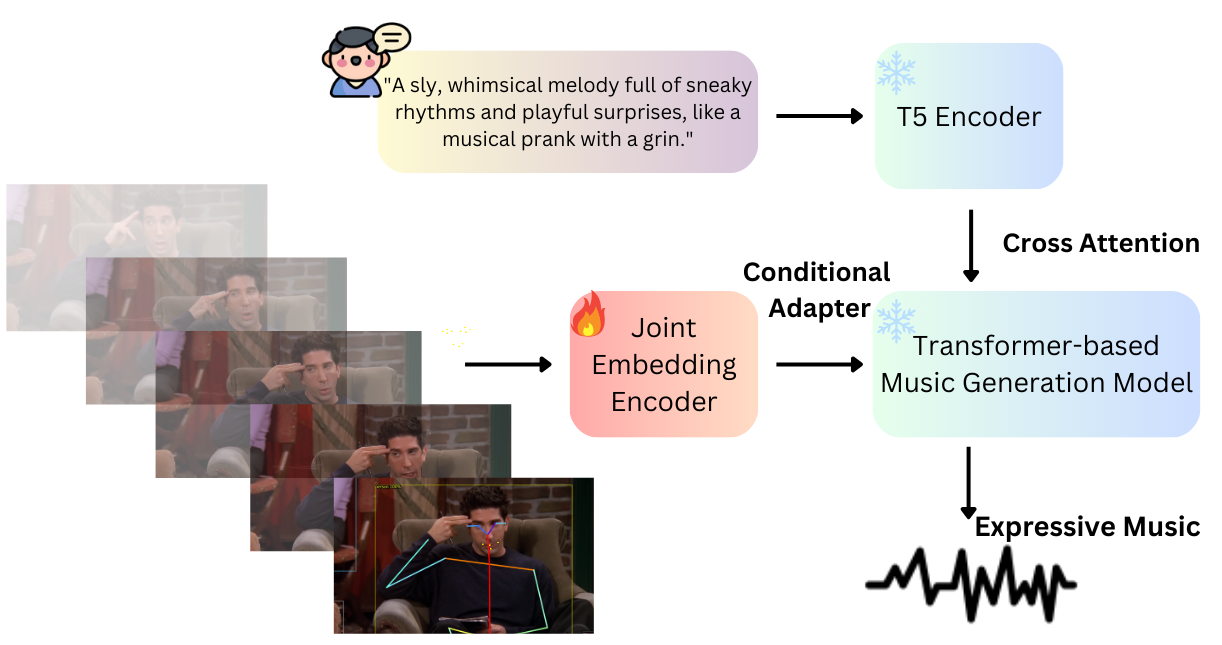}
    \caption{Overview of Expotion's multimodal inference pipeline, showing how visual gestures and facial expressions guide expressive music generation.}
    \label{fig:demo}
\end{figure}

Inspired by the idea that gestures and facial expressions can act as important guides for music, similar to what conductors do, we further explore \textit{visual} controls in this study and propose \textbf{Expotion}, a deep music generative model with multimodal controls--facial \textbf{exp}ression and upper body m\textbf{otion}, as well as text prompts. Expotion is designed to synthesize high-quality music that is both expressive (ensuring that the musical content reflects the emotional and expressive cues of the face and gestures) and temporally accurate (so that every change in motion and expression is accurately mirrored in the music) with the input video, as shown in Figure \ref{fig:demo}.

While some studies have explored audio–music generation tasks given videos—ranging from audio effects generation (e.g., \cite{mmaudio2024,difffoley2024,v_aura2024,foleycrafter2024,frieren2024,mmldm2024,visualechoes2024,videofoley2024,liu2024vatt,mo2024t2av,Du2023FoleyAnalogies}) video music background generation (e.g., \cite{muvi2024,vmas2024,video2music2024,Su2024V2Meow}), and dance music generation (e.g., \cite{Liang2024DanceComposer,Zhu2023Contrastive,Yu2023LORIS} among others), our work focuses on the subtle dynamics of gestures and facial expressions, emphasizing their fine-grained temporal synchronization with music generation. This opens up potential applications in real-time, interactive audiovisual systems.

To achieve our goal, we first introduce a newly curated dataset comprising of \textbf{7 hours of carefully synchronized video-music pairs} featuring expressive gestures and facial expressions closely matched to the corresponding music. Given the limited amount of data, we employ parameter-efficient fine-tuning \textbf{(PEFT) on a transformer-based text-to-music generation model}\cite{copet2024simplecontrollablemusicgeneration}—leveraging powerful ability of the model pretrained on a massive music-text data that have been shown to be effective in incorporating additional modalities \cite{lin2024contentbasedcontrolsmusiclarge,flores2024sketch2sound,liu2024vatt,mo2024t2av,guzhov2022audioclip,raffel2020exploring}. By fine-tuning only 4\% of the original model’s parameters \cite{lin2024contentbasedcontrolsmusiclarge}, our method seamlessly integrates multimodal visual inputs (facial expressions and upper-body gestures) using only 130 video–audio pairs for training, thereby minimizing network complexity while ensuring robust multimodal fusion. We also propose \textbf{an approach called temporal smoothing to ensure precise and efficient temporal alignment} between audio and video modalities.

Our experiments show that Expotion can generate high-quality, temporally accurate music that faithfully reflects the expressive nuances inherent in the visual inputs and textual descriptions. Although textual descriptions supply the primary contextual cues—leveraging the model’s strong text-understanding capabilities to guarantee baseline music quality—the addition of visual input further enhances alignment, expressiveness, and consistency. In comprehensive subjective and objective evaluations, Expotion consistently outperforms current state-of-the-art video-to-music generation models \cite{tian2024vidmusesimplevideotomusicgeneration} and multimodal captioning baselines across multiple metrics, including (1) Quality of Generated Music, (2) Beats and Tempo Consistency, (3) Text-Audio Similarity, and (4) Video–Music Consistency.

To the best of our knowledge, this work is the first to leverage synchronized expressive gestures and facial expressions for music generation. Our experiments show that incorporating visual features as control signals not only enhances the temporal alignment between the video and generated music but also improves text adherence and overall musical quality—highlighting the complementary strengths of both modalities. We believe that Expotion will empower artists with a more expressive, controllable, and interactive approach to music creation.

\begin{figure*}[ht!]
    \centering
    \includegraphics[width=0.9\linewidth]{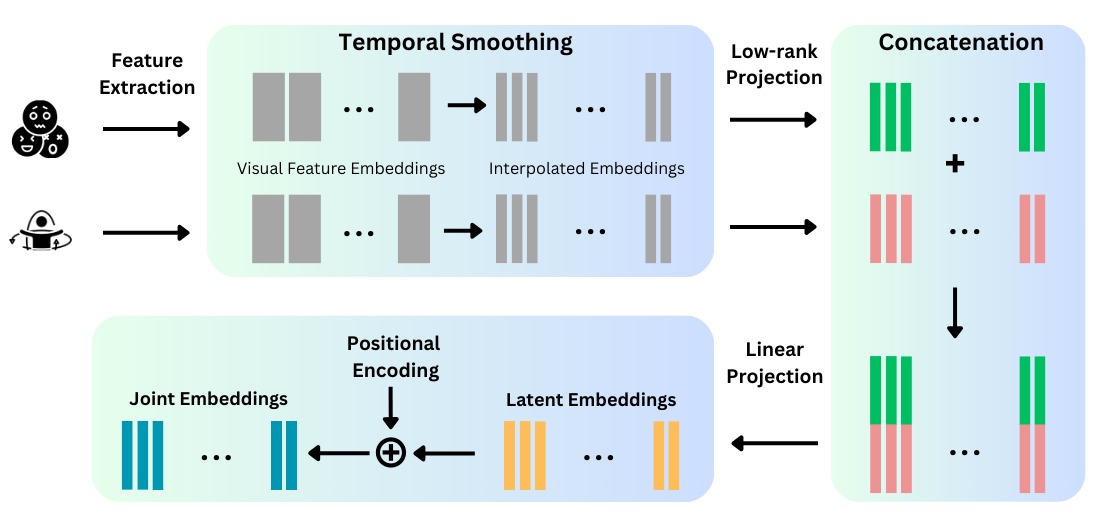}
    \caption{Joint embedding of visual features. Facial expression and motion features are first extracted from the given video and temporally smoothed through interpolation, producing refined intermediate representations. These temporally aligned embeddings then undergo dimensionality reduction through the low-rank projection layer. Then, the concatenated embeddings are projected to the same dimension as the MusicGen hidden layers. Finally, the positional encoding is added to the latent embeddings to form the final joint embeddings.}
    \label{fig:embeddings}
\end{figure*}

\section{Related Work}\label{sec:related_work}
We review three key paradigms for controllable music generation: visual and motion‐based control, textual and symbolic conditioning, and training and adaptation strategies.
\\

\noindent\textbf{Visual and Motion-Based Control}
Early interactive systems mapped facial or bodily features directly to sound. Valenti et al.’s \textit{Sonify Your Face} modulated audio via Bayesian classification of facial motion units \cite{valenti2010}, and Clay et al. translated whole-body emotional expressions into electronic-music parameters \cite{clay2012}. D2MNet extracted global style and local beat vectors from LMA-derived movement signals to drive an autoregressive generator \cite{huang2024d2mnet}. DeepTunes \cite{deeptunes2022} and \cite{huang2022icid} combine CNN‐based emotion detection with GPT‐2 lyric models, LSTMs, and transformers to jointly predict discrete and valence–arousal emotions and produce synchronized music (and lyrics) that closely reflect the user’s image input. Video-conditioned models such as VidMuse \cite{tian2024vidmusesimplevideotomusicgeneration}, V2Meow \cite{Su2024V2Meow}, and Video2Music \cite{video2music2024} align music with motion cues and scene context, while Foley-style systems translate individual events to sound via latent diffusion or transformers \cite{difffoley2024,videofoley2024,mmaudio2024}.

\noindent\textbf{Textual and Symbolic Conditioning}
Text-to-music models like MusicLM \cite{musiclm2023} and MusicGen \cite{copet2024simplecontrollablemusicgeneration} generate high-quality audio from descriptive prompts but lack explicit time-varying control. MusicGen-Melody extends this by conditioning on a reference melody track for pitch contours \cite{copet2024simplecontrollablemusicgeneration}. CoCoMulla leverages chord charts and drum patterns to control harmony and rhythm \cite{lin2024contentbasedcontrolsmusiclarge}, and Sketch2Sound introduces continuous vocal–imitation curves (loudness, brightness, pitch) alongside text to modulate audio generation \cite{flores2024sketch2sound}.

\noindent\textbf{Training and Adaptation Strategies}
Many video-to-audio and music models are trained from scratch on large paired datasets—such as AudioSet \cite{gemmeke2017audioset} and VGGSound \cite{chen2020vggsound}—to learn cross-modal correspondences \cite{mmaudio2024,difffoley2024,v_aura2024,foleycrafter2024,tian2024vidmusesimplevideotomusicgeneration}. When data is limited, parameter-efficient fine-tuning (PEFT) is effective: Sketch2Sound fine-tunes a single linear layer per control signal on a frozen diffusion backbone \cite{flores2024sketch2sound}, while CoCoMulla and AirGen attach lightweight adapters to MusicGen, tuning under 4\% of parameters for chord and rhythm control or music inpainting \cite{lin2024contentbasedcontrolsmusiclarge,lin2024arrange}. Expotion adopts a similar PEFT approach with visual multimodal inputs.

\section{Methodology}

Our approach consists of of 1) a joint embedding encoder to integrate temporally aligned video-based controls, and 2) a condition adaptor to fine-tune MusicGen by incorporating the learned joint visual embeddings. We froze the parameters of the vanilla Musicgen during training to preserve its text understanding ability.

\subsection{Joint Visual Embeddings}
To effectively incorporate facial expression and upper-body movement features from video, we adopt a joint embedding framework combining these two types of features, as shown in Figure \ref{fig:embeddings}. Since videos capturing facial expressions and upper-body gestures typically exhibit less dynamic motion, a relatively low frame rate is sufficient for smooth human perception. To address the discrepancy between this frame rate and the frame rate of MusicGen, we introduce targeted sampling strategies for video feature processing that differ from those applied to audio. We also propose an approach called temporal smoothing to ensure precise and efficient temporal alignment between audio and video modalities, maintaining the expressivity of video, and enhancing the overall multimodal integration. 

\subsubsection{Facial Expression Embedding}
To extract facial expression features from video, we employed MARLIN, a self-supervised learning framework specifically designed to derive universal facial representations from unannotated video data \cite{cai2023marlinmaskedautoencoderfacial}. For each frame, MARLIN generates features that capture information from both the current frame and its neighboring frames. To temporally align these facial expression features with the audio codes produced by Encodec in MusicGen, we utilized a specialized temporal smoothing strategy: first resampling the video from 30 fps to 80 fps and then, since MARLIN processes 16 frames simultaneously, obtaining the facial expression features in a frame rate of 5 fps by setting the stride to 16 frames. To minimize information loss from this downsampling, we applied linear interpolation to the facial expression feature $z_f \in \mathbb{R}^{T \times 768}$, where T is the total number of frames and 768 is the hidden dimension.  $z_f^{(i)} \in \mathbb{R}^{d_1}$ represents the feature at i-th frame. For a desired (possibly non-integer) time index \( t \), let 
\[
i = \lfloor t \rfloor, \quad \alpha = t - i,
\]
so that \( t \) lies between the \( i \)-th and \( (i+1) \)-th frames.
The interpolated feature \( \hat{z}_f^{(t)} \) is computed as
\begin{equation}
\hat{z}_{f_t} = (1-\alpha) \, z_{f_i} + \alpha \, z_{f_{i+1}}.
\end{equation}
This formula linearly weights the neighboring features, ensuring a smooth transition between frames.

We further compress the interpolated facial features by projecting them onto a low-dimensional space with dimension $d_1$ using a trainable matrix $W_f \in \mathbb{R}^{d_1\times768}$:
\begin{equation}\label{eq:low-rank1}
    {z'}_{f_i} = W_f^T \hat{z}_{f_i}  \in \mathbb{R}^{d_1}.
\end{equation}

\subsubsection{Motion Embedding}
We compare two motion representations: one extracted from the Synchformer visual encoder \cite{synchformer2024iashin} and another from RAFT optical flow \cite{teed2020raft}.

For Synchformer, we standardize videos fps and segment each video into 16-frame clips with a stride of 5, yielding outputs of shape \((T, 8, \text{dim})\), where the 8 dimension captures local temporal context. We flatten the first two dimensions, perform temporal interpolation as
\begin{equation}
\hat{z}_{m_t} = (1-\alpha)\,z_{m_i} + \alpha\,z_{m_{i+1}},
\end{equation}
and then project the interpolated features to a lower-dimensional space using a trainable matrix \(W_m \in \mathbb{R}^{d_2 \times D}\) (\(D\) = original feature dimension), yielding
\begin{equation}\label{eq:low-rank2}
z'_{m_i} = W_m^T\,\hat{z}_{m_i} \in \mathbb{R}^{d_2}.
\end{equation}

For RAFT, we sample the video at 5 fps to obtain frames \(\{I_t\}_{t=1}^{T}\) and compute the dense optical flow \(F_t \in \mathbb{R}^{H \times W \times 2}\) for each consecutive pair \((I_t, I_{t+1})\) using RAFT \cite{teed2020raft}. Each flow field is then processed by a Flow Embedding CNN \(\mathcal{F}\) to yield a compact feature vector:
\begin{equation}
z^{\text{flow}}_t = \mathcal{F}(F_t) \in \mathbb{R}^{256}.
\end{equation}

\(\mathcal{F}\) is composed of several convolutional layers with kernel size 3 and ReLU activations, followed by an adaptive average pooling operation. We perform temporal interpolation and project the resulting sequence to a lower-dimensional space in the same way as Sychformer. This ensures that both motion representations are aligned in time and compatible with the MusicGen transformer’s input requirements.

\subsubsection{Positional Embeddings}
We define a learnable matrix \( W_e \in \mathbb{R}^{(d_1+d_2) \times d} \) to fuse the embeddings mentioned above, together with a learnable positional embedding \( z_{\text{pos}, i} \in \mathbb{R}^{d_1+d_2} \) to support sequential modeling. The combined joint symbolic and acoustic embedding is computed as:
\[
z_i = W_e^T\Bigl( z_{f_i}; z_{m_i}] + z_{\text{pos}, i} \Bigr) \in \mathbb{R}^{d}.
\tag{6}
\]
Let \( T \) denote the total number of frames. Then, the overall sequential joint embedding is given by:
\[
z = \{z_1, z_2, \dots, z_T\} \in \mathbb{R}^{T \times d}.
\tag{7}
\]

\subsection{Condition Adaptor}
Adopting a similar approach to CoCoMulla~\cite{lin2024contentbasedcontrolsmusiclarge}, we extend the idea of a condition adaptor to handle time-varying video inputs, such as facial expressions and motion. In a standard Transformer, each self-attention layer processes \( T \) hidden embeddings (one per Encodec frame). In our approach, the final \( L \) layers of the MusicGen decoder expand this to \( 2T \) embeddings by adding \( T \) \emph{condition prefix} positions, which encode the control information. Specifically, we insert a sequence of learnable input embeddings into the \((N - L + 1)\)th decoder layer to initiate the condition prefix. In this prefix, the hidden states go only through self-attention layers (omitting cross-attention). Let
\[
H^p_l \in \mathbb{R}^{T \times d} \quad (N - L + 1 \leq l \leq N)
\]
be the output for the condition prefix, with \( H^p_0 \) as the learnable input embeddings. The condition prefix is computed as:
\[
Q^p_l,\, K^p_l,\, V^p_l = \text{QKV-projector}\bigl(H^p_l + Z_l\bigr),
\]
\[
H^p_{l+1} = \text{Self-Attention}\bigl(Q^p_l,\, K^p_l,\, V^p_l\bigr),
\]
where \( Z_l \) are the sequential joint embeddings (defined in Eq.~(7)). No causal mask is applied here, and the condition prefix does not attend to the Encodec tokens.

For the remaining part, the hidden states \( H_l \in \mathbb{R}^{T \times d} \) (for \( 1 \leq l \leq N \)) are processed normally. Their standard attention output \( S_l \) is computed as:
\[
Q_l,\, K_l,\, V_l = \text{QKV-projector}(H_l),
\]
\[
S_l = \text{Self-Attention}\bigl(Q_l,\, K_l,\, V_l\bigr).
\]

To incorporate condition information in the last \( L \) layers, we compute cross attention \( S'_l \) between \( Q_l \) and \(\{K^p_l, V^p_l\}\) using self-attention, fusing \( Q_l \) with \( Q^p_l \):
\[
S'_l = \text{Self-Attention}\bigl(Q_l + Q^p_l,\, K^p_l,\, V^p_l\bigr).
\]
A learnable gating factor \( g_l \) (initialized to zero) combines the outputs:
\[
H_{l+1} = \text{Cross-Attention}\bigl(S_l + g_l \cdot S'_l,\, \text{text}\bigr).
\]

In our implementation, all MusicGen layers (including QKV-projector, Self-Attention, and Cross-Attention) are frozen; only \( H^p_0 \), \( W_p \), \( W_a \), \( W_e \), \( z_{\text{pos}} \), and \( g_l \) are trainable.

\section{Experiments}
\subsection{Dataset}
Due to the lack of sufficient paired video-audio data with clear facial features, we curated our own dataset by collecting the data manually. We recruited volunteers to record their facial expressions and upper body movements while listening to 30-second audio clips. Before starting the recording, the volunteers were asked to listen to the music track once, allowing them time to think about the facial expressions and body movements that would align with the music. The audio clips used were licensed instrumental tracks from Epidemic Sound, ensuring no vocals were present. The collection includes a variety of music genres, such as pop, jazz, blues, classical, and epic, among others. We were able to collect 7 hours of paired video-audio data. The paired video-audio data are then chopped into 10 seconds per clip.  We set aside 30 minutes of data for test and validation. We also generated captions for each audio clip with audio-captioning model SALMONN \cite{tang2024salmonngenerichearingabilities} to be used as text prompts during training and inference. The prompt given to SALMONN for captioning is 'Please describe the music'. In the data collecting process, volunteers gave informed consent, agreeing that their videos would be used only for research and anonymized to protect their privacy.

\subsection{Implementation}
Our base model, MusicGen (text-only), consists of three main parts: a pre-trained EnCodec, a pre-trained T5 encoder, and an acoustic transformer decoder. The decoder has 48 layers, each with causal self-attention and cross-attention to process text prompts. MusicGen uses EnCodec, a Residual Vector Quantization (RVQ) auto-encoder~\cite{défossez2022highfidelityneuralaudio}, to convert audio sampled at 32,000 Hz into discrete codes at 50 Hz, which are then passed to the transformer decoder.

We trained the proposed model using four A1000 GPUs, employing an initial learning rate of 1e-02 and a batch size of 10 consisting of ten 10-second audio samples, for 40 epochs. Training was stopped after 40 epochs to prevent overfitting. During training, the model's parameters were updated using a cross-entropy reconstruction loss. In the low-rank projection step, we set $d_1$ in Equation ~\eqref{eq:low-rank1} and $d_2$ in Equation ~\eqref{eq:low-rank2} to be 12.

\begin{table*}[htbp]
\centering
\hspace*{-0.5cm}
\footnotesize
\setlength{\tabcolsep}{1pt}
\resizebox{\textwidth}{!}{%
\begin{tabular}{lccc ccc cc cc}
\toprule
\textbf{} & \textbf{} & \textbf{} 
& \multicolumn{3}{c}{\textbf{General Music Quality}} 
& \multicolumn{2}{c}{\textbf{Rhythm Alignment}} 
& \multicolumn{2}{c}{\textbf{Text/Video-Audio Consistency}} \\
\cmidrule(lr){4-6} \cmidrule(lr){7-8} \cmidrule(lr){9-10}
\textbf{}  & \textbf{Motion Feat.} & \textbf{Prompt}  
& \textbf{FAD-VGG}\(\textcolor{green}{\downarrow}\) 
& \textbf{Avg. KL}\(\textcolor{green}{\downarrow}\) 
& \textbf{Avg. IS Score}\(\textcolor{red}{\uparrow}\) 
& \textbf{Tempo Err. (bpm)}\(\textcolor{green}{\downarrow}\) 
& \textbf{F1-score (Beat)}\(\textcolor{red}{\uparrow}\) 
& \textbf{Text-Audio}\(\textcolor{red}{\uparrow}\) 
& \textbf{Video-Audio}\(\textcolor{red}{\uparrow}\) \\
\midrule
face        & --         & Generated & 2.52  & 0.74  & 1.45  & 31.53  & \textbf{0.38} & 0.48 & 0.55 \\
motion      & RAFT       & Generic   & 3.56  & 1.05  & 1.16  & 28.07  & 0.37 & 0.38 & \textbf{0.65} \\
motion      & RAFT       & Generated & 2.25  & 0.66  & 1.53  & 33.71  & 0.37 & 0.52 & 0.59 \\
motion      & Syncformer & Generic   & 3.52  & 1.08  & 1.20  & 31.03  & 0.35 & 0.37 & 0.61 \\
motion      & Syncformer & Generated & \textbf{1.93}  & \textbf{0.65}  & \textbf{1.57}  & 32.89  & 0.36 & 0.52 & 0.61 \\
face+motion & Syncformer & Generated & 2.55  & 0.67  & 1.49  & 32.72  & 0.37 & \textbf{0.54} & 0.59 \\
MusicGen    & --         & Generated & 2.76  & 0.79  & 1.54  & 35.84  & 0.33 & 0.50 & 0.42 \\
Video2Music & --         & --        & 18.97 & 1.32  & 1.01  & \textbf{26.13}  & 0.38 & 0.25 & 0.59 \\
VidMuse     & --         & --        & 9.91  & 1.10  & 1.33  & 33.01  & 0.37 & 0.36 & 0.52 \\
\bottomrule
\end{tabular}%
}
\caption{Comparison of music quality, rhythm alignment, and text/video-audio consistency metrics across different configurations.}
\label{tab:combined_metrics}
\end{table*}

\subsection{Baselines}

Since our model addresses a novel domain for which no existing opensource models are specifically trained with both video (facial expression and motion) and text controls, we selected vanilla MusicGen (text-conditioned) and two recent video-conditioned music generation models—VidMuse\cite{tian2024vidmusesimplevideotomusicgeneration} and Video2Music\cite{Kang_2024}—as our baselines. Video2Music uses an Affective Multimodal Transformer to generate emotionally aligned, expressive symbolic music (chords) from video inputs by leveraging semantic, motion, scene, and emotion features, while VidMuse integrates both local and global visual cues through a Long-Short-Term Visual Module.

\subsection{Evaluation}
Our evaluation consists of two parts: subjective and objective evaluations. Because MusicGen only accepts textual input, we generate fused multimodal captions from the video–audio pairs to serve as prompts for the text-only MusicGen baseline. This approach ensures a fair comparison by providing equivalent descriptive information across all models.

\subsubsection{Objective Evaluation}
In our objective evaluation, we assess the music generated from three perspectives: (1) the inherent quality, (2) rhythm alignment, and (3) text/video-audio consistency. To measure the inherent quality of the music, we employed Frechet Audio Distance(FAD) with VGGish embeddings which measures perceptual similarity to real music\cite{kilgour2019frechet, hershey2017cnn}, Kullback-Leibler (KL) divergence which quantifies distributional alignment with real audio labels\cite{kong2020panns, koutini2021passt}, and Inter-Sample Score (IS Score) which captures the diversity among generated samples\cite{salimans2016improved}. To evaluate the rhythm alignment between generated and ground truth music, we calculated the tempo error, which is defined as how far an estimated BPM(beats-per-minute) deviates from the true (ground-truth) BPM, and beat consistency between the generated and reference music. To evaluate the text/video-audio consistency, we use CLAP\cite{wu2024largescalecontrastivelanguageaudiopretraining} and LanguageBind\cite{zhu2024languagebindextendingvideolanguagepretraining} respectively. We employ these models trained with contrastive learning approach to measure model's ability to generate music that reflects the semantic meaning of the text and video. 

\subsubsection{Subjective Evaluation}
We gathered participants of varying level of musical background to rate the generated music—across various configurations and the baseline—over five groups of six videos. They were shown the text prompts without being informed which model produced each track. Ratings were based on the following criteria:

\begin{itemize}[noitemsep, topsep=0pt, leftmargin=*]
\item \textbf{Musicality:} How effectively the audio captures key musical qualities.
\item \textbf{Text-audio Similarity:} The degree of alignment between the generated music and the provided textual prompt.
\item \textbf{Video-audio Consistency:} The extent to which the music corresponds with the video content in terms of tempo and emotional expression.
\item \textbf{Creativity:} The uniqueness and innovativeness of the generated audio.
\end{itemize}

\subsection{Ablation studies}
We conducted ablation studies to evaluate the effects of various experimental setups. Specifically, these studies explored the following model setups: (1) the use of different motion features (RAFT versus Synchformer) and (2) the use of generic prompts of `music with catchy melody' versus detailed prompts generated by the audio-captioning model.

\section{Results}

\subsection{Objective Evaluations}

\textbf{General Music Quality.} The results in Table \ref{tab:combined_metrics} demonstrate that models incorporating motion information, particularly those using the Syncformer features with generated prompts, consistently outperform others across general music quality metrics, especially the baselines. This configuration performs best overall, with the lowest FAD and KL scores and the highest IS Score, indicating realistic, diverse, and well-aligned music generation. Comparatively, models using the RAFT features perform less consistently, and those using only facial input or generic prompts yield weaker results. Baseline methods like VidMuse and Video2Music show significantly poorer FD and KL scores, highlighting the advantage of multimodal control of our model. 

\noindent\textbf{Rhythm Alignment.} Among all the configuration tested, the model trained with RAFT motion features and generic captions achieves the lowest Average Tempo Error (28.07 BPM), indicating better temporal alignment with the audio compared to other proposed methods. Video2Music achieves the lowest tempo error because it transcribes the audio into MIDI and computes rhythmic characteristics in the form of note density and loudness from the audio--a proxies for the music's rhythms \cite{video2music2024}. The baseline model MusicGen shows the poorest performance in all tempo and beat tracking metrics, underscoring its non-expressiveness in controlling generated music.

\noindent\textbf{Text/Video-Audio Consistency} 
The text-audio similarity scores reveal that multimodal conditioning (face and motion) significantly enhance text-alignment compared to text-only conditioning baseline (MusicGen), although they do not provide explicit textual information. The face+motion model achieves the highest text-music scores (0.54).In contrast, video-audio similarity scores show that models trained with generic (non-descriptive) captions achieve stronger alignment with video (e.g., 0.65 and 0.61), suggesting that text conditioning may not benefit—and may even hinder—video-audio consistency. The poor video-audio alignment of MusicGen outputs and the low text-audio alignment in music generated by Video2Music and VidMuse is justied as these models are not conditioned on the respective modalities.

\subsection{Subjective Evaluation}
\begin{figure}[h!]
    \centering
    \includegraphics[width=1\linewidth]{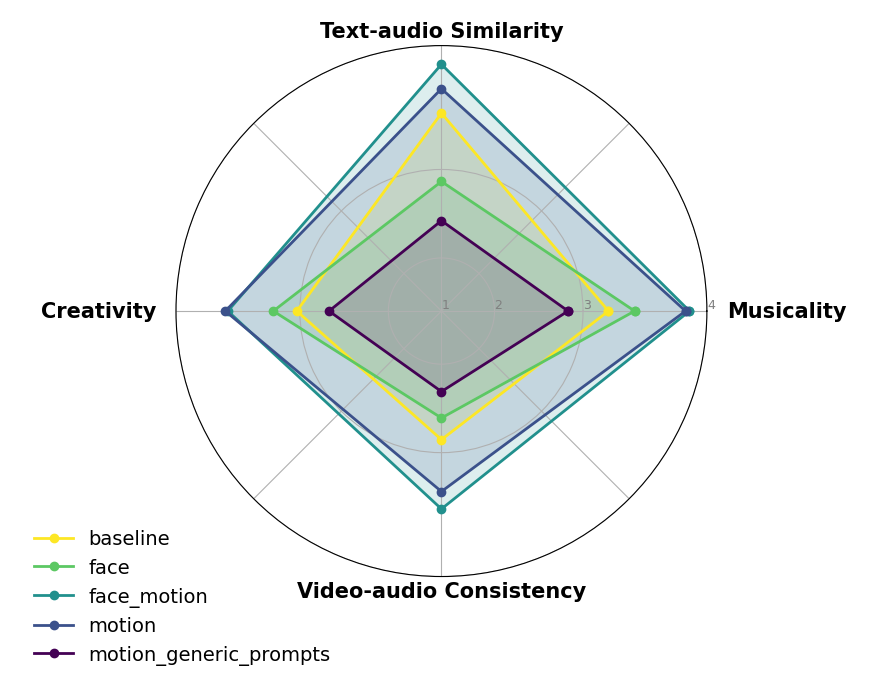}
    \caption{Subjective Evaluation Results on Four Metrics}
    \label{fig:sub}
\end{figure}

Figure \ref{fig:sub} presents a subjective analysis of the quality of music generated by five models—vanilla MusicGen(baseline) and four variants of our models trained with different configurations. A total of 13 participants (6 female, 7 male) aged 18–40 years (Median = 27) took part. Based on self‐reported musical training, 20 $\%$ of participants have beginner level of experience in music, 20 $\%$ intermediate, and 60 $\%$ professional.

Notably, all of our models—except for one—outperform the baseline in creativity and musicality, suggesting that incorporating facial expressions and motion cues enables the system to better capture the expressive qualities of the input video. This expressiveness is reflected in the generated music, which participants perceive as more musical and creative. Our model trained with only motion features and generic prompts, perform poorly in all evaluation metrics, indicating that generic textual prompts are insufficient to guide high-quality music generation. Without meaningful textual context, visual features alone do not provide enough semantic grounding to produce good music. The superior performance of motion features over facial features in both objective and subjective evaluations likely stems from the nature of the dataset: participants found it easier to convey musical cues through movements rather than facial expressions, resulting in richer and more expressive motion data.
These findings highlight that visual and textual modalities are complementary: while textual input provides semantic intent, visual features—especially motion—enrich the generated music’s expressiveness. Overall, our model, integrating features from both modality is capable for producing music that is coherent, expressive, and creative as perceived by human. 
\subsection{Ablation Studies}
\subsubsection{RAFT vs. Synchformer}
The comparison between RAFT and Syncformer as motion feature extractors reveals notable differences in both general music quality and rhythm-related metrics as shown in Table \ref{tab:combined_metrics}. Syncformer outperforms RAFT across FAD, and IS Score, indicating more realistic and slightly more diverse music generation. However, there is no significant difference between the choice of these two motion featrues in terms of tempo error and beat accuracy of the generated music. These finding suggests that Syncformer may capture more expressive and semantically rich motion patterns, whereas RAFT may be better at preserving rhythmic consistency in simpler contexts. 

\subsubsection{Generic vs. Generated Captions}
Comparing models trained with generated captions to those using generic captions, we observed that although music generated with generic captions yielded lower CLAP scores—likely due to receiving minimal textual information—they achieved better tempo accuracy than those trained with generated prompts (tempo error of 28.07 BPM tempo versus 33.71 BPM when comparing models trained with same configuration except for choice for prompts). This suggests that adding extra textual context may introduce noise or distract from the purely visual motion cues, ultimately reducing temporal accuracy.

\section{Conclusion}
Expotion demonstrates that visual cues—specifically, body movements and facial expressions—can effectively serve as expressive controls for music generation. By leveraging a pretrained text-to-music model \cite{copet2024simplecontrollablemusicgeneration} and applying parameter-efficient fine-tuning, our approach achieves notable improvements from the original text-only conditioning MusicGen using only 130 clips (6 hours) of training data in 40 epochs. Integrating these multimodal signals with textual prompts, Expotion produces music that shows strength in musicality, creativity, and temporal accuracy, as evidenced by enhanced beat, tempo, and overall semantic consistency across text, video, and audio. A temporal smoothing strategy further ensures fine-grained alignment between the visual cues and the generated music. Our results outperform state-of-the-art baselines, and subjective studies confirm that the combination of facial and motion features yields superior performance, while objective evaluations highlight that motion features—particularly those extracted via Synchformer—strike an optimal balance between rhythmic consistency and expressive dynamics. Overall, Expotion represents a promising step toward more expressive, controllable, and interactive audiovisual music generation systems.
\bibliography{ISMIRtemplate}
\end{document}